\documentclass[journal=ancac3, articletitle=true]{achemso}
\setkeys{acs}{articletitle = true}
\setkeys{acs}{maxauthors = 10}

\usepackage[T1]{fontenc}
\usepackage[english]{babel}
\usepackage{graphicx}
\usepackage{amssymb}
\usepackage{amsmath}
\usepackage{amsfonts}
\usepackage{color}
\usepackage{bm}

\newfloat{scheme}{htbp}{los}
\floatname{scheme}{Scheme}
\floatname{chart}{Chart}
\newfloat{graph}{htbp}{loh}

\usepackage{chemformula} 
\usepackage[version = 4]{mhchem} 

\author{Kazutoshi Masuda}
\affiliation{Department of Basic Science, Graduate School of Arts and Sciences, The University of Tokyo, Komaba 3-8-1, Meguro, Tokyo 153-8902, Japan}

\author{Takahiro Yokoyama}
\affiliation{Leibniz-Institut für Polymerforschung Dresden e.V., Hohe Straße 6, 01069 Dresden, Germany}
\alsoaffiliation{Institut für Theoretische Physik, Technische Universität Dresden, 01069 Dresden,Germany}

\author{Miho Yanagisawa}
\email{myanagisawa@g.ecc.u-tokyo.ac.jp}
\affiliation{Department of Basic Science, Graduate School of Arts and Sciences, The University of Tokyo, Komaba 3-8-1, Meguro, Tokyo 153-8902, Japan}
\alsoaffiliation{Komaba Institute for Science, Graduate School of Arts and Sciences, The University of Tokyo, Komaba 3-8-1, Meguro, Tokyo 153-8902, Japan}
\alsoaffiliation{Department of Physics, Graduate School of Science, The University of Tokyo, Komaba 3-8-1, Meguro, Tokyo 153-8902, Japan}
\alsoaffiliation{Research Center for Complex Systems Biology, Universal Biology Institute, The University of Tokyo, Komaba 3-8-1, Meguro, Tokyo 153-8902, Japan}

\author{Arash Nikoubashman}
\email{anikouba@ipfdd.de}
\affiliation{Leibniz-Institut für Polymerforschung Dresden e.V., Hohe Straße 6, 01069 Dresden, Germany}
\alsoaffiliation{Institut für Theoretische Physik, Technische Universität Dresden, 01069 Dresden,Germany}

\title{Interfacial Packing of DNA Nanostars Regulates Dynamics on Synthetic Cell Membranes}

\date{\today}

\begin{document}

\maketitle

\begin{abstract}
DNA nanostructures are emerging as programmable components for engineering synthetic cell membranes, yet how their collective packing and deformability regulate molecular dynamics at membrane interfaces remains poorly understood. Here, we investigate the packing and mobility of star-shaped DNA nanostructures (nanostars) with tunable stiffness on lipid-coated droplets. The negatively charged nanostars spontaneously adsorb onto cationic membranes, and their interfacial packing is changed by the bulk DNA concentration and droplet size, which together determine the number of encapsulated nanostars. Combining fluorescence recovery after photobleaching experiments with coarse-grained molecular dynamics simulations, we reveal distinct packing–dynamics relationships for rigid and soft nanostars. For rigid nanostars, diffusion first decreases gradually and then drops sharply with increasing interfacial packing, approaching a dynamically arrested state consistent with jamming-like behavior. In contrast, at equivalent experimental conditions, soft nanostars systematically reach lower interfacial packing fractions and show a weaker decrease in apparent mobility. This behavior is consistent with their weaker membrane affinity, which facilitates adsorption--desorption with the bulk. When this exchange is suppressed in simulations, crowding reduces the lateral diffusion of both nanostar types, but produces distinct dense configurations: soft nanostars become strongly deformed, whereas rigid nanostars largely retain their shape and form interlocked, gear-like arrangements. Nanostar adsorption also impedes lipid diffusion, while differences in membrane affinity result in distinct lipid mobile fractions. These findings reveal how the interplay between nanostar packing and deformability regulates molecular transport at membrane interfaces, providing a physical design principle for tuning lateral fluidity and crowding in artificial cells.
\end{abstract}

\vspace{10mm}
Aqueous droplets surrounded by lipid membranes, including liposomes and lipid-coated droplets, serve as versatile platforms for synthetic biology and artificial cell engineering. \cite{podolsky2021, yanagisawa2022, yanagisawa2025} Their functionality critically depends on membrane properties such as stability and fluidity, which are governed by interfacial organization. To tune these properties, polymers and nanoparticles are often adsorbed at membrane interfaces.\cite{simovic2012, umar2022} While such colloidal additives can stabilize the interface, high interfacial coverage can suppress membrane fluidity and permeability, both of which are essential for emulating cell-like behavior.\cite{ramadurai2009lateral, javanainen2012, houser2016} Understanding how interfacial packing regulates membrane dynamics while preserving interfacial stability therefore remains a major challenge.

In two-dimensional colloidal systems, increasing particle density can lead to caging and ultimately jamming, strongly suppressing particle mobility.~\cite{o2003jamming} 
Similar crowding-induced slowing of diffusion has been reported for rigid particles and macromolecules associated with supported lipid membranes.\cite{horton2010, liu2023effects} Yet it remains unclear whether such packing--dynamics relationships apply to deformable soft particles coupled to closed, fluid membranes. Unlike planar substrates, membrane-coated droplets provide a finite, curved, and laterally fluid interface on which adsorbed particles and lipids can rearrange together. Moreover, in a closed compartment, the number of encapsulated particles relative to the available membrane area depends on the droplet size, even at a fixed bulk concentration, which presents a challenge when comparing droplets of different sizes, but also provides a direct means to tune interfacial packing.

DNA nanostructures provide an ideal materials platform for studying and utilizing such packing--dynamics relations: their geometry, anisotropy, and conformational rigidity can be programmed with nanometer precision, providing a versatile route to engineer particles with controlled shape and deformability.\cite{hong2017, lee2024} Recent studies have shown that, like double-stranded DNA of several kbp or more,\cite{maier2000, herold2010dna} nanometer-sized DNA nanostructures can spontaneously adsorb onto cationic lipid membranes, including planar membranes, liposomes, and membrane-coated droplets.\cite{langecker2014, zhan2023, samanta2024, fang2026, masuda2026} 
Membrane association of DNA nanostructures is also strongly modulated by the ionic environment, particularly by divalent cations.\cite{morzy2021}
The number of phosphate groups and the conformational rigidity of DNA can strongly influence membrane adsorption, with larger and stiffer structures generally adsorbing more strongly.\cite{morzy2023} Previous experiments with DNA nanostructures on supported lipid membranes found density-dependent packing transitions, such as long-range ordered structures for Y-shaped DNA nanostructures.\cite{avakyan2017, dong2020} These observations suggest that the interplay between membrane adsorption, rigidity, and interfacial packing of DNA nanostructures may jointly regulate their mobility and the dynamics of the surrounding lipids.

Here, we test this hypothesis by systematically investigating how interfacial packing of DNA nanostructures governs dynamics on synthetic cell membranes, using star-shaped DNA nanostructures (hereafter termed ``nanostars'') as model soft colloids. By encapsulating nanostars within cationic lipid-coated droplets at varying droplet radii and bulk nanostar concentrations, we control the number of enclosed nanostars and thereby tune their interfacial packing. Combining fluorescence measurements with coarse-grained molecular dynamics (MD) simulations, we identify a packing-controlled crossover in nanostar mobility that strongly depends on nanostar stiffness. We further show that lipid diffusion is suppressed by nanostar interfacial crowding, where differences in membrane affinity lead to distinct lipid mobile fractions for rigid and soft nanostars. These findings establish DNA nanostars as programmable soft colloids whose collective packing governs both their own mobility and the dynamics of the surrounding synthetic-cell membrane.

\section{Results and discussion}
\subsection{Interactions of individual DNA nanostars with lipid membrane}
\begin{figure}[htb]
\includegraphics[width=7.5cm] {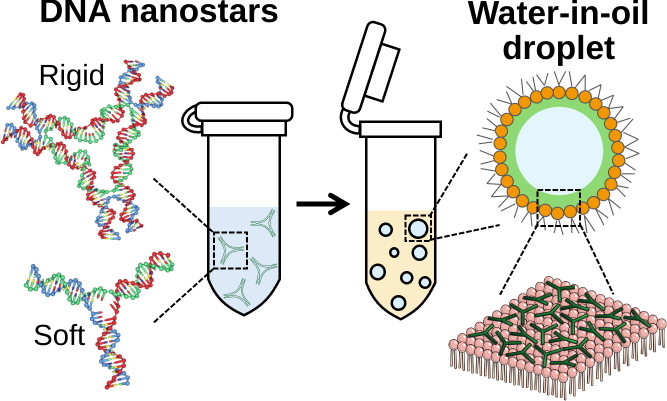}
\caption{ 
Representative conformations of rigid and soft stars obtained from coarse-grained MD simulations, and schematic illustration of a DOTAP-coated droplet encapsulating DNA nanostars. 
}
\label{fig:fig1}
\end{figure}

To investigate how particle deformability influences interfacial packing and dynamics, we designed two types of three-armed DNA nanostars that differ in their conformational rigidity (Fig.~\ref{fig:fig1}); rigid nanostars contain $N_\text{nuc}=267$ nucleotides in a double-duplex configuration, whereas soft stars contain $N_\text{nuc} = 102$ nucleotides in a single-duplex structure (see Fig.~S1 for sequence information).
Unlike conventional Y-shaped DNA motifs that may form networks through intermolecular hybridization,\cite{masuda2026, dong2020, sato2020, sato2024, kurokawa2017} the DNA nanostars used here lack sticky ends and therefore behave as isolated nanoparticles.
We estimated the molecular footprint, i.e., the instantaneous area occupied by a DNA star, from MD simulations in the dilute limit, yielding $S_0 = 82\, \mathrm{nm^2}$  and $S_0 = 20\, \mathrm{nm^2}$ for rigid and soft nanostars, respectively (see Sec. S7 in the Supporting information). The approximately fourfold difference in $S_{0}$, despite rigid nanostars containing only 2.6 times as many nucleotides as soft nanostars, indicates a more loosely packed spatial arrangement of nucleotides in the rigid nanostars.

In principle, positively charged lipid membranes attract DNA nanostructures through electrostatic interactions with their negatively charged sugar-phosphate backbone, with stronger adsorption at higher fractions of cationic lipids.\cite{maier2000} However, previous experiments found that $\lambda$-DNA chains desorbed from lipid membranes containing only 10\% cationic 1,2-dioleoyl-3-trimethylammonium-propane (DOTAP) by mass when the NaCl concentration exceeded $50\,\text{mM}$.\cite{maier2000}
Therefore, we used a synthetic cell membrane composed of a 100 mol\% DOTAP monolayer to maximize electrostatic attraction between the membrane and the DNA nanostars.

To estimate the adsorption strength, we modeled the interaction between individual DNA nanostars and the lipid membrane as a combination of screened electrostatic attraction and short-range hydration repulsion.\cite{cherstvy2014} Other contributions to the adsorption free energy, including those associated with counterion release and the entropic cost of water reorganization, were neglected for simplicity.
The hydration repulsion, which is commonly employed for hydration interactions between lipid and hydrophilic interfaces,\cite{rand1989, pera2004} arises from the energetic cost associated with perturbing the interfacial hydration layers between two surfaces. It is described by an exponentially decaying pressure, $P_{\mathrm{hyd}}(h)=P_0\exp\left(-h/\lambda_{\mathrm{hyd}}\right),$ where $h$ is the surface-to-surface separation between the DNA nanostar and the membrane, $P_0 = 10^9\, {\rm Pa}$ is the hydration pressure extrapolated to contact,\cite{simon1991, mukhina2019} and $\lambda_{\mathrm{hyd}}=0.25\, \mathrm{nm}$ is the characteristic decay length of the hydration interaction.\cite{marrink1993} 
The corresponding hydration free energy was calculated as $G_{\mathrm{hyd}}(h) = S_0\int_h^\infty P_{\mathrm{hyd}}(z)\,\text{d}z$.

To estimate the electrostatic contributions to the adsorption free energy, the nanostars were approximated as uniformly charged disks with area $S_0$ and effective charge $Q=f_\text{DNA}N_\text{nuc}$, where $f_\text{DNA} \leq 1$ regulates the effective fraction of the nominal DNA charge. The lipid membrane was represented as an uniformly charged planar surface, with surface charge density\cite{zhao2012, gao:aaps:2013} $1.4\,\text{e/nm}^2$ scaled by an effective charge fraction $f_\text{mem} \leq 1$. Exploiting the rotational symmetry of this disk--plane configuration, we solved the non-linear Poisson--Boltzmann (PB) equation in a two-dimensional axisymmetric geometry. Because both the DNA nanostars and the cationic DOTAP membrane possess high nominal charge densities, their effective charges in solution are expected to be lower than their bare charges. 
To bracket the physically plausible range of electrostatic interactions, we considered two limiting charge conditions in the PB calculations. 
To estimate the upper bound, we assumed that both the DNA phosphate groups and DOTAP molecules carry their full nominal charges, so that $f_\text{DNA}=f_\text{mem}=1$. 
For the lower bound, the DNA effective charge fraction was reduced to $f_\text{DNA}=0.1$ to account for Manning counterion condensation,\cite{manning1969} whereas the membrane effective charge fraction was set to $f_\text{mem}=0.2$, chosen to be consistent with the experimentally measured $\zeta$-potential of approximately $40\,\text{mV}$ for DOTAP membranes.\cite{miatmoko2023} Note that these lower $f$ values should be regarded as estimates, since Manning theory assumes an idealized line charge, while the exact conversion from $\zeta$-potential to the membrane surface potential requires knowledge of the slip-plane position.\cite{delgado2007}

\begin{figure}[htb]
\includegraphics[width=7.5cm] {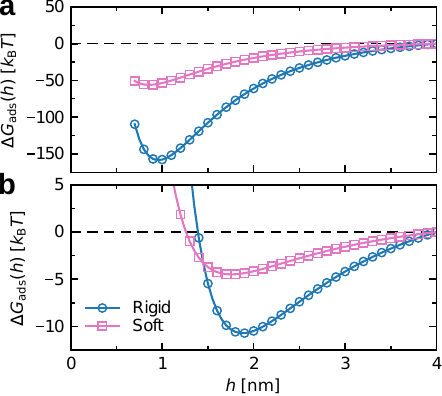}
\caption{Adsorption free energy of DNA nanostars at the membrane, $\Delta G_{\mathrm{ads}}$, shown for the (a) upper-bound condition ($f_{\mathrm{DNA}}=f_{\mathrm{mem}}=1.0$) and (b) lower-bound condition ($f_{\mathrm{DNA}}=0.1$ and $f_{\mathrm{mem}}=0.2$) . Blue and pink curves denote rigid and soft nanostars, respectively.}
\label{fig:fig2_delta_G}
\end{figure}

For the upper-bound estimate, the adsorption free energy, $\Delta G_\text{ads}$, is minimized at a distance of about $h \approx 0.9\,\text{nm}$ from the membrane, with $\Delta G_\text{ads} \approx -55\,k_\text{B}T$ and $-160\,k_\text{B}T$ for the soft and rigid nanostars, respectively, indicating effectively irreversible adsorption (Fig.~\ref{fig:fig2_delta_G}a). For the lower-bound estimate, the minimum of $\Delta G_\text{ads}$ shifts to roughly $h \approx 1.7\,\text{nm}$, with substantially weaker adsorption free energies of approximately $-4.5\,k_\text{B}T$ and $-11\,k_\text{B}T$ for the soft and rigid nanostars, respectively  (Fig.~\ref{fig:fig2_delta_G}b). 
Thus, the soft nanostars are only weakly bound in the lower-bound model, suggesting reversible and highly dynamic membrane association. In contrast, the more negative free energy of the rigid nanostars indicates substantially stronger binding, and hence less frequent adsorption--desorption events.

\subsection{Adsorption and packing of encapsulated DNA nanostars}
\begin{figure*}[htbp]
\includegraphics[width=1.0 \linewidth] {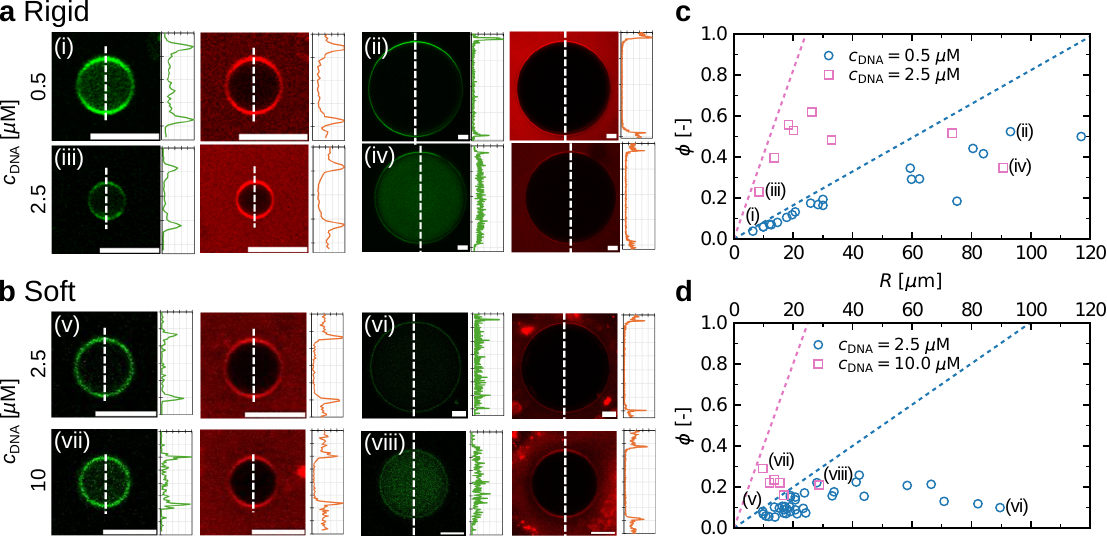}
\caption{(a,b) Confocal cross-section images showing nanostars (green) inside droplets covered with a lipid membrane (red), for (a) rigid and (b) soft DNA nanostars at various DNA concentrations, $c_\text{DNA}$, and droplet radii, $R$, as indicated. Fluorescence intensity profiles along the dashed lines are shown beside each image. Scale bars indicate $20\,\mu\text{m}$. (c,d) The effective interfacial packing fraction $\phi$ of (c) rigid and (d) soft nanostars, estimated from confocal fluorescence images, as a function of droplet radius $R$. Different colors indicate different $c_\text{DNA}$, while dashed lines indicate the ideal case, $\phi_\text{id}$, in which all encapsulated nanostars are localized at the membrane interface ($I_\mathrm{r}=1$ in Eq.~\eqref{eq:phi}).}
\label{fig:fig3_partition_DNA}
\end{figure*}

In the experiments, we encapsulated the nanostars within cell-sized droplets coated with DOTAP and observed fluorescently labeled DNA nanostars by confocal laser scanning microscopy over a broad range of droplet radii, ranging from $R=5\,\mu\text{m}$ to $120\,\mu\text{m}$ (Fig.~\ref{fig:fig3_partition_DNA}). 
Because the molecular footprint $S_{0}$ of rigid nanostars is approximately four times larger than that of soft nanostars, different DNA concentration ranges were used for the two systems. Figures~\ref{fig:fig3_partition_DNA}a,b show selected representative cases (small droplets with $R \approx 7\,\mu\text{m}$ and large droplets with $R \approx 90\,\mu\text{m}$) at two DNA concentrations ($c_{\mathrm{DNA}} = 0.5~\mu\text{M}$ and $2.5\,\mu\text{M}$ for rigid nanostars, and $c_{\mathrm{DNA}} = 2.5~\mu\text{M}$ and $10\,\mu\text{M}$ for soft nanostars). 
The total number of nanostars in a droplet scaled with its volume; hence, the number of stars available for membrane adsorption was determined by $c_{\mathrm{DNA}}$ and $R$. 

To quantify the effective interfacial packing fraction $\phi$ of DNA nanostars from the confocal microscopy data, we determined the relative fluorescence intensity $I_\text{r}$ of the membrane-adjacent region (see Fig.~S2; Sections S1 and S2). Because the first adsorbed layer cannot be distinguished from subsequent wetting layers within the optical resolution of confocal laser scanning microscopy, $I_\text{r}$ reflects the relative nanostar concentration near the membrane rather than direct adsorption (Fig.~S3). Assuming that the DNA concentration inside the droplet is equal to the initial bulk concentration $c_\text{DNA}$ and neglecting the droplet's curvature, 
\begin{equation}
    \phi \equiv \frac{I_\mathrm{r} c_\mathrm{DNA} N_\mathrm{A} R S_0}{3} ,
    \label{eq:phi}
\end{equation}
with Avogadro constant $N_\text{A}$. To determine whether the estimated interfacial packing is limited by the number of available nanostars within the droplet, we compared $\phi$ with the idealized value $\phi_\text{id}$ obtained by setting $I_\text{r} =1$ in Eq.~\eqref{eq:phi}, which implies that all nanostars are adsorbed onto the membrane. 

For both rigid and soft nanostars, $\phi$ generally increased with increasing droplet radius, $R$, and DNA concentration, $c_\text{DNA}$, as expected from the increasing number of nanostars per unit membrane area (Fig.~\ref{fig:fig3_partition_DNA}c,d). For rigid nanostars, $\phi$ closely followed the predicted linear dependence, but increasingly fell below $\phi_\text{id}$ as $R$ increased. At sufficiently large $R$, $\phi$ saturated at approximately $\phi \approx 0.4$ for $c_{\mathrm{DNA}}=0.5\,\mu\mathrm{M}$ and $\phi \approx 0.6$ for $c_{\mathrm{DNA}}=2.5\,\mu\mathrm{M}$ (Fig.~\ref{fig:fig3_partition_DNA}c). The soft nanostars likewise exhibited an initial increase in $\phi$, but the linear regime extended over a much smaller range of $R$. Further, the estimated $\phi$ values were significantly more scattered and reached a smaller plateau value of $\phi \approx 0.2$, far below the ideal limit $\phi_\text{id}$ (Fig.~\ref{fig:fig3_partition_DNA}d).

\begin{figure*}[htbp]
\includegraphics[width=1.0 \linewidth] {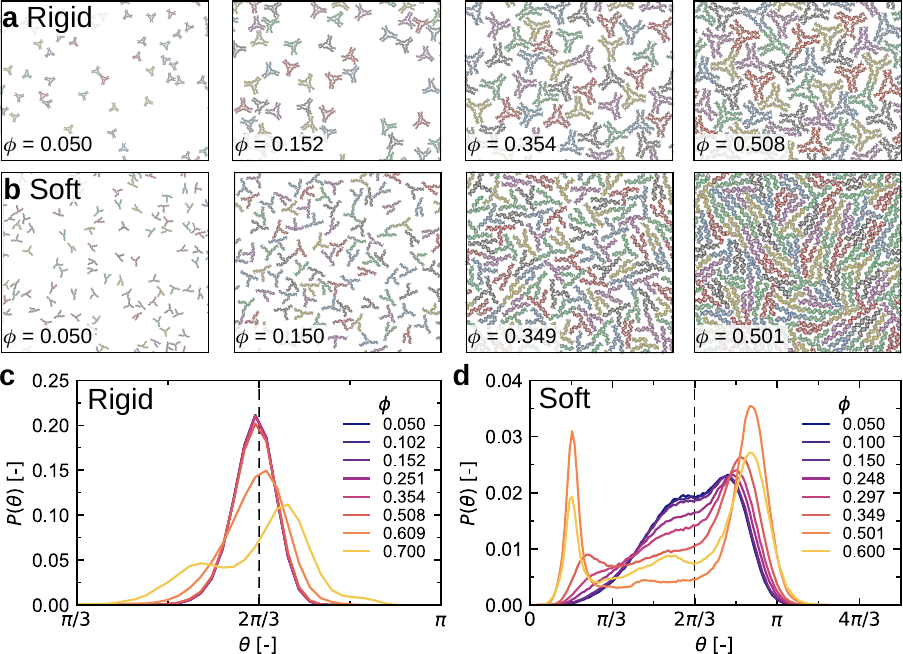}
\caption{Representative MD simulation snapshots of (a) rigid and (b) soft DNA nanostars at different packing fractions $\phi$, as indicated. The colors represent the individual stars. (c,d) Probability distribution of inter-arm angle, $P(\theta)$, of (c) rigid and (d) soft nanostars at different $\phi$.}
\label{fig:fig4_phi_dependence}
\end{figure*}

To understand the molecular origin responsible for this relatively sparse interfacial localization, we performed MD simulations of DNA nanostars adsorbed to a smooth, flat substrate.
In the simulations, nanostars were confined to the surface by a short-range attraction along the surface normal, which ensured complete adsorption throughout the simulations. The number of nanostars, $N_\text{star}$, was kept constant, and the concentration was varied by adjusting the lateral size of the simulation box. 
The packing fraction $\phi$ from the simulations is then defined as $\phi=N_\text{star}S_0/\left( L_x L_y \right)$ , with $L_x$ and $L_y$ denoting the dimensions of the simulation box.
Figures~\ref{fig:fig4_phi_dependence}a,b show representative simulation snapshots of rigid and soft nanostars, respectively, at different packing fractions $\phi$. Notably, rigid-star configurations appear more sparse at the same $\phi$, owing to their less compact double-duplex architecture which has a $35\,\%$ lower nucleotide density than the soft stars.

These simulation snapshots already reveal increasing deformation of the nanostars with increasing $\phi$ relative to the dilute limit, with the effect being particularly pronounced for the soft nanostars. We quantified this deformation via the distribution of angles between neighboring arms, $\theta$, computed from the scalar product of the unit vectors connecting the center of mass of a star to the tip of an arm. In the dilute limit ($\phi \approx 0.01$), the resulting probability distribution $P(\theta)$ exhibits a sharp peak at $\theta = 2\pi/3$ for rigid stars, whereas soft stars display a much broader distribution spanning $\sim \pi/3$ to $\sim \pi$ (Figs.~\ref{fig:fig4_phi_dependence} c,d), confirming their substantially greater conformational flexibility. As $\phi$ increased, $P(\theta)$ broadened for both nanostar types, indicating increased conformational distortion due to intermolecular crowding. Importantly, these changes emerged at packing fractions close to the experimentally observed saturation values for each motif, namely $\phi \approx 0.6$ for rigid nanostars, and $\phi \approx 0.2$ for soft nanostars (Figs.~\ref{fig:fig3_partition_DNA}c,d). These results suggest that the experimentally observed saturation of interfacial localization is strongly tied to the significant distortion of the nanostars, consistent with an increasing free-energy penalty for accommodating additional nanostars at the membrane.

\subsection{Diffusivity of DNA Nanostars}
\begin{figure}[htbp]
\centering
\includegraphics[width=1.0\linewidth]{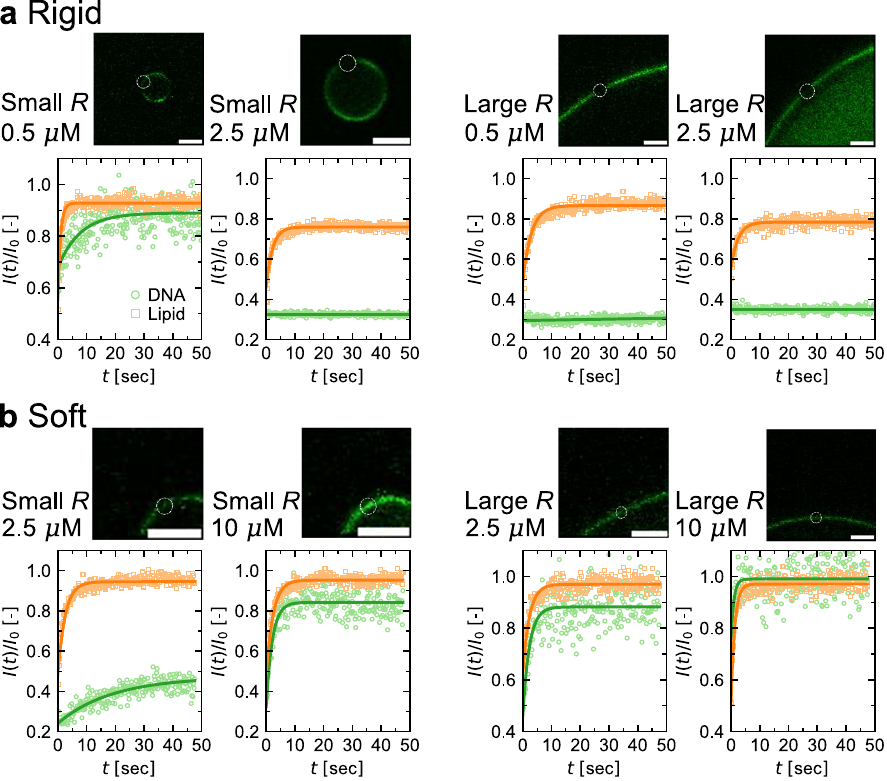}
\caption{
(a) Representative FRAP curves for nanostars (green circles) and lipids (orange squares) for small and large droplets. The darker-colored curves are the fitted results using Eq.~\eqref{eq:frap}. $I_0$ experimentally denotes the average fluorescence intensity of the ten pre-bleaching frames. Photobleached regions are indicated by dashed white circles in the images. Scale bars are 10 $\mu$m.
The DNA concentrations are $c_{\mathrm{DNA}} =$ 0.5 or 2.5$\,\mu\text{M}$ for rigid stars and $c_\text{DNA}=2.5$ or $10 \mu M$ for soft stars.}
\label{fig:fig5_frap}
\end{figure}

We now turn to the dynamic consequences of interfacial crowding, which may restrict the lateral mobility of membrane-bound nanostars.
To quantify these effects, we performed fluorescence recovery after photobleaching (FRAP) measurements on individual droplets to determine the diffusion coefficients ($D$) and mobile fractions ($M_{\mathrm{f}}$) of both nanostar types (Fig.~S5). Circular regions (radius $w \approx 2\,\mu\text{m}$) on the membrane were photobleached, and the fluorescence recovery was monitored over time (Figs.~\ref{fig:fig5_frap}a, b).
Recovery curves were obtained under all conditions and were well described by a single-exponential function with recovery time $\tau_\mathrm{rec}$ (see Materials and Methods section):
\begin{equation}\label{eq:frap}
I(t) = I_0 + (I_{\infty} - I_0)\, \left( 1 - e^{-t/\tau_\mathrm{rec}} \right) ,
\end{equation}
from which the diffusion coefficients $D$ were calculated as
\begin{equation}\label{eq:diffusion}
D = \frac{w^{2}}{4\tau_\mathrm{rec}} .
\end{equation}
Control experiments showed that unbound nanostars in the droplet interior exhibited complete fluorescence recovery (Fig.~S4), confirming that the limited recovery observed at the membrane is associated with membrane binding rather than photobleaching artifacts.

\begin{figure}[htb]
\centering
\includegraphics[width=1\linewidth]{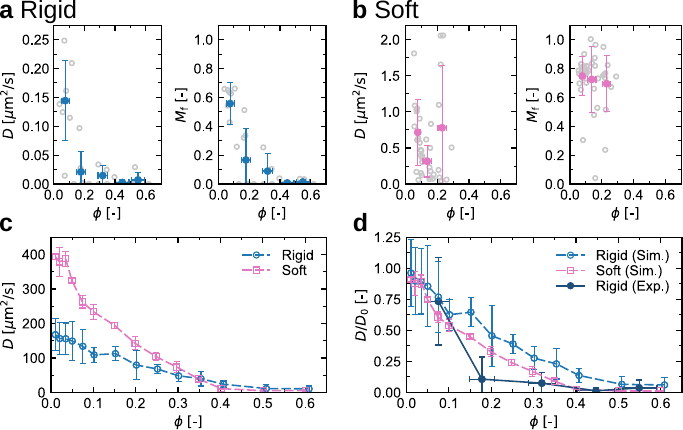}
\caption{(a, b) Experimentally obtained diffusion coefficient, $D$, and mobility fraction, $M_{\mathrm{f}}$, of (a) rigid and (b) soft DNA nanostars as functions of packing fraction, $\phi$. Open symbols show original data, while filled symbols show averaged values from linearly spaced bins; error bars denote the standard deviation.
(c) $D$ from simulations of rigid and soft nanostars as functions of $\phi$. (d) $D$, normalized by its value at infinite dilute, $D_0$, as functions of $\phi$, from simulations and experiments, as indicated.
}
\label{fig:fig6_dna_diffusion}
\end{figure}

The extracted values of $D$ and $M_{\mathrm{f}}$ reveal that interfacial packing strongly regulates the nanostar mobility, but the trends differed markedly between rigid and soft stars (Figs.~\ref{fig:fig6_dna_diffusion}a,b). To visualize the overall trends, we grouped individual measurements (open symbols) into linearly spaced bins in $\phi$ (filled symbols).
The diffusion coefficient, $D$, and mobile fraction, $M_{\mathrm{f}}$, of rigid stars decreased markedly with increasing $\phi$, indicating a gradual transition into a dynamically arrested state (Fig.~\ref{fig:fig6_dna_diffusion}a). The concurrent reduction in $D$ and $M_{\mathrm{f}}$ further suggests that crowding restricts not only the translational diffusion of individual stars but also the molecular exchange within the membrane-associated nanostar layer. Dynamic arrest is approached at $\phi \gtrsim 0.3$, well below the jamming threshold of hard disks at $\phi \approx 0.84$ \cite{}; this onset is closer to the jamming range of $\phi \approx 0.48 - 0.7$ reported for quasi-2D experiments on cross-shaped granular materials,\cite{zheng2017, stannarius2022} suggesting that shape anisotropy causes arrest at lower packing fractions.

In contrast, $D$ and $M_\text{f}$ of soft stars varied only weakly across the experimentally accessible $\phi$-range (Fig.~\ref{fig:fig6_dna_diffusion}b). Notably, $D$ exhibited substantial variability near the highest packing fraction, $\phi \sim 0.2$. This weaker $\phi$-response may reflect their deformability, which allows local rearrangement under crowding. Further, reversible membrane binding of soft stars could facilitate transient desorption and re-adsorption events, which would contribute to fluorescence recovery in addition to lateral diffusion along the membrane. To test whether the exchange of bleached membrane-bound nanostars with unbleached nanostars from the bulk increases the apparent mobility of the nanostars, we compared FRAP measurements performed at two different positions on the droplet membrane (Fig.~S6). Bleaching at the top surface provides greater access to the surrounding bulk phase, whereas bleaching in the cross-sectional membrane region restricts bulk access while preserving lateral diffusion along the membrane. For soft stars, the mobile fraction was significantly higher at the top surface than in the cross-sectional geometry ($p=0.003$, paired $t$-test), consistent with an additional recovery pathway through nanostar exchange with the bulk region. In contrast, FRAP measurements for the rigid stars showed no detectable difference between the two bleached regions ($p=0.972$), indicating that their fluorescence recovery is dominated by lateral motion within the membrane-associated layer. These measurements therefore support bulk-mediated membrane exchange as an additional contribution to the apparent mobility of soft, but not rigid, nanostars.

To isolate the effect of nanostar stiffness on lateral diffusion, we used a sufficiently strong nanostar--surface attraction in our MD simulations to suppress desorption events.
Thus, both rigid and soft stars remained adsorbed and underwent two-dimensional diffusion confined to the surface. Their diffusion coefficient $D$ was calculated directly from the mean-square displacement (MSD) of individual nanostars (Fig.~S8). 
At low packing fractions ($\phi \to 0$), the soft stars had an approximately $2.6$ times higher diffusion coefficient $D_0$, consistent with their correspondingly lower number of nucleotides, $N_\text{nuc}$, compared to the rigid nanostars. A similar inverse scaling $D_0 \propto N_\text{nuc}^{-1}$ was observed in previous experiments on $\lambda$-phage DNA adsorbed to planar lipid membranes.\cite{maier1999, maier2000} As the packing fraction increased, $D$ decreased almost linearly with $\phi$ for both rigid and soft nanostars, showing that crowding suppresses their intrinsic lateral diffusion irrespective of stiffness; the weak $\phi$-dependence of $D$ measured experimentally for soft nanostars is therefore consistent with an additional contribution from bulk-mediated exchange.

\begin{figure}[htbp]
\centering
\includegraphics[width=14cm]{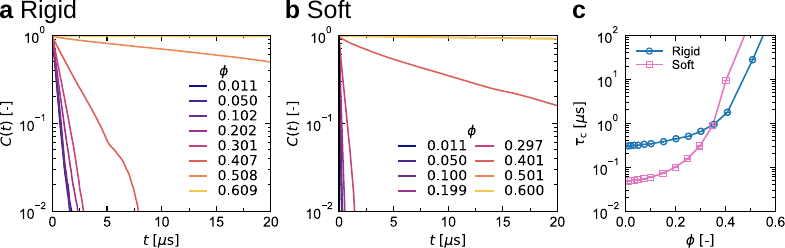}
\caption{(a,b) The arm vector autocorrelation function $C(t)$ of (a) rigid and (b) soft stars at different $\phi$. (c) The characteristic relaxation time $\tau_c$, obtained from exponential fits of $C(t)$, as a function of $\phi$.
}
\label{fig:fig7_BACF}
\end{figure}

Notably, the normalized diffusion coefficient, $D/D_0$, decayed significantly faster for soft stars and reached $D/D_0 \approx 0$ at a slightly lower packing fraction ($\phi \approx 0.4$) compared to rigid stars ($\phi \approx 0.5$). To elucidate the molecular origin of this difference, we examined the internal relaxation of the nanostars via the autocorrelation function of the angle $\theta$ between two arms
\begin{equation}
    C(t)=\langle \cos[\theta(t_0)] \cos[\theta(t_0 + t)] \rangle .
\end{equation}
For both rigid and soft stars, $C(t)$ decays exponentially (Figs.~\ref{fig:fig7_BACF}a,b) and we extracted the characteristic internal relaxation time, $\tau_\text{c}$, by fitting $C(t)  =\exp{(-t/\tau_c)}$. 
For dilute systems, $\tau_\text{c}$ increases only gradually with $\phi$, but then rises sharply as crowding restricts arm relaxation (Fig.~\ref{fig:fig7_BACF}c). This slowing of internal relaxation is more pronounced for soft stars, whose $\tau_\text{c}$ surpasses that of rigid stars at $\phi \gtrsim 0.35$. This crossover occurs near the earlier onset of translational slowdown (Fig.~\ref{fig:fig7_BACF}c), linking their translational arrest to the suppression of internal relaxation. The difference in internal relaxation is also evident in the simulation snapshots (Figs.~\ref{fig:fig4_phi_dependence}f,g): rigid stars largely retain their shape and form an interlocked gear-like arrangement, whereas soft stars become strongly deformed to maximize interfacial packing.

\subsection{Suppressed diffusion of lipids by adsorbed DNA nanostars}
\begin{figure}[htb]
\centering
\includegraphics[width=7.5cm]{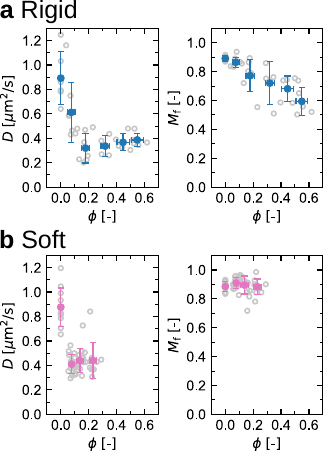}
\caption{(a, b) Experimentally obtained $D$ and $M_{\mathrm{f}}$ of lipids as a function of the star area fraction, $\phi$, for membranes decorated with (a) rigid and (b) soft stars. Symbols and error bars represent mean values and standard deviations, respectively. Faint open symbols indicate the original experimental data.
}
\label{fig:fig8_lipid_diffusion}
\end{figure}

Previous studies of protein-rich membranes have revealed a decrease in lateral mobility  of both proteins and surrounding lipids.\cite{javanainen2012, jeon2016} Therefore, we examined whether the diffusion coefficient, $D$, and mobile fraction ,$M_{\mathrm{f}}$, of the lipids decrease with increasing nanostar packing fraction, $\phi$, and whether this response differs between rigid and soft nanostars.
To test this hypothesis, 1 in 1000 membrane lipids was replaced with a dye-labeled analog (Rho-PE), while a small fraction of nanostars was likewise fluorescently labeled (see SXX for the details).
We determined $D$ and $M_{\mathrm{f}}$ from FRAP experiments (orange lines in Fig.~\ref{fig:fig5_frap}), and plotted the resulting values against $\phi$ in Fig.~\ref{fig:fig8_lipid_diffusion}. 
Note that the experimentally accessible range of $\phi$ for soft stars ($\phi \lesssim 0.25$) is considerably smaller than that of rigid stars ($\phi \lesssim 0.55$), as discussed above.

To establish a baseline, we first characterized the lipid dynamics of bare membranes ($\phi = 0$), finding $D \approx 0.9\,\mu\text{m}^2/\text{s}$ and $M_\text{f} \approx 0.9$.
For membranes decorated with rigid nanostars, $D$ gradually decreased with increasing $\phi$ and reached a plateau of $\approx 0.4\,\mathrm{\mu m^2/s}$ at $\phi \gtrsim 0.2$ (Fig.~\ref{fig:fig8_lipid_diffusion}a, left). Concurrently, the mobile fraction $M_{\mathrm{f}}$ decreased almost linearly from $M_\text{f} \approx 0.9$ to $\approx 0.6$ as the nanostar packing fraction increased to $\phi \approx 0.55$ (Fig.~\ref{fig:fig8_lipid_diffusion}a, right).
For membranes with soft nanostars, lipid diffusivity decreased sharply by $\phi \sim 0.1$ and remained approximately constant at $D \approx 0.4\,\mu\text{m}^2/\text{s}$ (Fig.~\ref{fig:fig8_lipid_diffusion}b, left). In contrast to the rigid nanostar case, most lipids remained highly mobile ($M_\text{f} \approx 0.9$) over the experimentally accessible $\phi$ range (Fig.~\ref{fig:fig8_lipid_diffusion}b, right).

The stronger suppression of lipid diffusivity for the soft nanostars may originate from their approximately 1.5-fold higher projected nucleotide density at the same nominal packing fraction $\phi$, potentially resulting in more DNA–lipid contacts per unit area.
Despite this different sensitivity, the similar plateau value of $D$ suggests a common limiting lipid mobility for sufficiently dense DNA layers. The distinct trends in $M_{\mathrm{f}}$ qualitatively mirror those observed for the nanostars themselves (Fig.~\ref{fig:fig6_dna_diffusion}), and thus likely reflect differences in the persistence of the DNA--membrane association (see Fig.~\ref{fig:fig2_delta_G} and corresponding discussion). Reversible adsorption--desorption of soft nanostars likely makes their interactions with nearby lipids more transient, allowing most lipids to redistribute during fluorescence recovery even though their diffusion is slowed. In contrast, rigid nanostars bind much more tightly to the membrane, thereby preventing the redistribution of nearby lipids over the FRAP timescale.

\section{Conclusions}
By combining experiments on membrane-coated droplets with coarse-grained molecular dynamics simulations, we show that interfacial packing and bending rigidity jointly regulate the mobility of adsorbed DNA nanostars and surrounding lipids. The closed droplet geometry couples bulk DNA concentration and droplet size to the number of encapsulated nanostars per unit membrane area, thereby providing direct control over interfacial crowding. Despite using membranes composed entirely of cationic lipids, the experimentally measured nanostar packing fractions remain systematically below the ideal limit of full adsorption. At a given droplet size and DNA concentration, soft nanostars reach significantly lower packing fractions than rigid nanostars, suggesting more frequent bulk-mediated desorption--adsorption exchange. These differences indicate incomplete interfacial localization and are consistent with a weaker effective electrostatic attraction of soft nanostars to the membrane.

FRAP experiments show that the diffusivity and mobile fraction of adsorbed rigid nanostars decrease simultaneously with increasing packing fraction, approaching a dynamically arrested state. In contrast, soft nanostars exhibit a weaker dependence of their apparent mobility on packing fraction, likely because bulk-mediated nanostar exchange provides an additional pathway for fluorescence recovery that partly masks the suppression of lateral diffusion. By excluding this exchange pathway in our simulations, we find that crowding suppresses the intrinsic lateral diffusion of both rigid and soft nanostars, albeit through distinct crowded configurations: rigid nanostars largely retain their shape and form interlocked, gear-like arrangements, whereas soft nanostars undergo substantial deformation and exhibit earlier suppression of internal relaxation.

Notably, nanostar adsorption also suppresses lipid mobility; soft nanostars cause a sharper decrease in lipid diffusivity, which we attribute to their 1.5-fold higher potential nucleotide--lipid contacts per unit membrane area at the same nominal packing fraction. In contrast, the mobile fraction of lipids qualitatively mirror those of the nanostars and reflect their different adsorption--desorption dynamics: the mobile fraction decreases with increasing packing fraction of rigid nanostars, but remains nearly constant for the more transiently adsorbed soft nanostars. Together, these results establish DNA nanostars as programmable components for tuning interfacial crowding and lateral membrane mobility through molecular architecture, offering a physical design principle for controlling membrane dynamics in synthetic cells.

\section{Materials and Methods}
\subsection{Materials}
1,2-dioleoyl-3-trimethylammonium-propane (DOTAP; chloride salt) was purchased from Avanti Polar Lipids. Lipids were dissolved in chloroform (Nacalai Tesque), and lipid films obtained after solvent evaporation were redissolved in hexadecane (Nacalai Tesque) as the oil phase. Sodium chloride (NaCl; FUJIFILM Wako Pure Chemical Co.) and Tris–HCl buffer (pH 8.0; Nacalai Tesque) were dissolved in ultrapure water (Invitrogen) to prepare the buffer solution for soft DNA nanostars. Tris-Acetate-EDTA buffer (pH 8.3; Nacalai Tesque) with Magnesium Acetate Tetrahydrate (Nacalai Tesque) added was used to prepare the buffer solution for rigid DNA nanostars. For DNA nanostars, lyophilized DNA oligonucleotides (Eurofins Genomics Japan) were dissolved in ultrapure water and added to the aqueous phase as appropriate. DNA stock solutions were prepared at 2 mM and stored at -20~$^\circ$C until use. All materials were used as received without further purification.

\subsection{DNA nanostars preparation}
We designed DNA oligomers based on previously reported architectures to assemble DNA nanostars.\cite{kurokawa2017, he2005, sun2009} At high temperatures, the DNA oligomers remain dissociated due to entropic stabilization, whereas upon cooling they hybridize via Watson–Crick base pairing to form DNA nanostars. Two DNA nanostar architectures were designed: a rigid nanostar with arms consisting of two double-stranded DNA (dsDNA) helices and a soft nanostar with arms consisting of a single dsDNA helix. DNA nanostars display a single melting transition at $T_{\mathrm{m}}$ and remain as isolated motifs. The melting temperatures were $T_{\mathrm{m}} = 65~^\circ$C for rigid and $69~^\circ$C for soft nanostars, respectively.

For self-assembly of rigid and soft nanostars, the constituent DNA oligomers were separately mixed at equimolar ratios and diluted in their respective buffers ($20\,\text{mM}$ Tris–HCl, pH 8.0, $350\,\text{mM}$ NaCl for rigid nanostars; $20\,\text{mM}$ Tris–HCl, pH 8.0, $350\,\text{mM}$ NaCl for soft nanostars). The solution was heated to 80~$^\circ$C for 10 minutes and then slowly cooled to 10~$^\circ$C at a rate of 0.01~$^\circ$C,s$^{-1}$ using a T-Gradient thermocycler (Biometra, Germany). For fluorescence analysis, one of the constituent DNA oligomers was labeled with 6-carboxyfluorescein (FAM) for the soft nanostar and fluorescein isothiocyanate (FITC) for the rigid nanostar.

\subsection{Droplet preparation}
Water-in-oil droplets encapsulated by lipid monolayers were prepared as follows. DOTAP was first dissolved in chloroform at a final concentration of 10 mM. An aliquot (50 $\mu$L) of the lipid solution was transferred to a Durham tube, and the solvent was evaporated under a gentle stream of nitrogen gas to form a dry lipid film at the bottom of the tube. Hexadecane ($500\,\mu\text{L}$) was then added, and the mixture was sonicated at approximately 60~$^\circ$C for 90 minutes to obtain a lipid-in-oil solution with a final lipid concentration of $1\,\text{mM}$. Approximately 0.1 mol\% of the total lipids was replaced with Rho-PE for fluorescence analysis. The solution was subsequently cooled slowly from 60~$^\circ$C to room temperature (approximately 25~$^\circ$C) with intermittent vortex mixing to ensure homogeneous lipid dissolution.

To prepare artificial cells with DNA nanostars localized beneath the membrane, $2\,\mu\text{L}$ of the aqueous solution containing nanostars was added to $40\,\mu\text{L}$ of the DOTAP-in-hexadecane solution after sonication. 
Here, to minimize the ion difference in the buffers, the mixture of soft nanostars are diluted with the buffer for rigid nanostars. Since both systems contained divalent cations in the buffer, differences in membrane–DNA electrostatic interactions were minimized. Throughout this study, the DNA nanostar concentration, $c_{\mathrm{DNA}}$, denotes the final concentration after dilution, which was assumed to be identical in the droplets.
The mixture was then gently tapped to induce droplet formation without bulk emulsification. The resulting droplets were transferred onto a silicone-coated glass-bottom dish (Matsunami) to prevent adhesion to the glass surface and were used for subsequent observations and measurements.

\subsection{Fluorescence intensity analysis}
Droplets with DNA motifs were observed by confocal laser scanning fluorescence microscopy (Olympus IX83 with FV1200; Olympus). Fluorescent images of FAM and FITC, which were excited by a mercury lamp or a laser ($473\,\text{nm}$ for DNA nanostars and $495\,\text{nm}$ for lipid membrane), were obtained using fluorescence filter sets (U-FBNA for mercury lamp; $470\,\text{nm}$ to $495\,\text{nm}$ for laser; Olympus). The pinhole size was fixed to be about $1\,\mu\text{m}$. The obtained images were analyzed by the National Institutes of Health ImageJ software.

\subsection{FRAP}
Fluorescence recovery after photobleaching (FRAP) experiments \cite{axelrod1976} were performed using the same confocal microscopy setup as described above.
Circular regions of interest with a radius of \(w \approx 2\,\mu\mathrm{m}\) were photobleached by applying a high-intensity laser pulse.
Fluorescence recovery was then monitored over time under the same imaging conditions as before photobleaching.
The fluorescence intensity in the bleached region was corrected for background fluorescence and normalized by the pre-bleach intensity.
The pre-bleach intensity, $I_{\mathrm{pre}}$, was defined as the average fluorescence intensity immediately before photobleaching.
The intensity immediately after photobleaching was denoted as $I_0$, and the plateau intensity after recovery was denoted as $I_{\infty}$.
The mobile fraction $M_{\mathrm{f}}$ was calculated as the fraction of fluorescence recovered relative to the fluorescence lost by photobleaching, as $M_{\mathrm{f}} =(I_{\infty}-I_0)/(I_{\mathrm{pre}}-I_0)$. 

\subsection{Simulation Model}
We employ the coarse-grained oxDNA2 model,\cite{poppleton:joss:2023, snodin:jcp:2015} where single-stranded DNA is represented as sequences of rigid-body nucleotides, interacting via effective bonded and non-bonded potentials.\cite{ouldridge:jcp:2011, sulc:jcp:2012}
All molecular dynamics (MD) simulations were conducted using the oxDNA package with GPU acceleration.\cite{rovigatti:jcc:2015}
All simulations were conducted in the $NVT$ ensemble with an implicit solvent, where $N$ is the number of nucleotides and $V$ is the volume of the simulation box.
The fundamental simulation units of length, mass and energy were set as $\sigma=0.8518\, \text{nm}$, $m = 5.24\times10^{-25} \, \text{kg}$, and $\varepsilon = 4.142 \times 10^{-20} \, \text{J}$, respectively.
The temperature was kept constant at $T=300\,\text{K}$ using a Langevin thermostat with translational friction coefficient $\gamma =m/\tau$ with $\tau=\sqrt{m \sigma^2/\varepsilon} \approx 3.03 \,\text{ps}$ being the intrinsic simulation timescale.
Electrostatic interactions between DNA chains are modeled using Debye-H{\"u}ckel theory, with the concentration of implicit monovalent salt chosen as $0.5\,\text{M}$.
The equations of motion are integrated using a time step of $\Delta t = 15.15\,\text{fs}$.

DNA nanostars were prepared manually using the oxView software\cite{poppleton:nar:2020, bohlin:natprotoc:2022} from the single-stranded DNA sequences (see Fig.~S1 in the SI).
We placed DNA nanostars on an attractive flat substrate, modeled using a standard Lennard-Jones potential acting on all nucleotides with interaction strength $\varepsilon_\text{sub} = k_\text{B}T$.
This potential acts only on the direction normal to the substrate, thus neglecting lateral friction from the substrate. In all simulations, the dimension normal to the substrate was fixed at $L_z=17.0\, \text{nm}$. The number of stars was kept constant as $N_\text{star} = 48$, and $N_\text{star} = 96$ for rigid and soft nanostars, respectively.
The simulation box dimensions were chosen as $L_{x}:L_{y}=9:4\sqrt{3}$ for rigid nanostar system and $L_{x}:L_{y}=2:\sqrt{3}$ for soft nanostar system.
To vary the area fraction $\phi$ of DNA nanostars, we incrementally applied two-dimensional deformation to the simulation box, compressing both $L_x$ and $L_y$ by $1\%$ at each step and then relaxing the system for $30.3\,\text{ns}$ before the next deformation. 
Before measuring nanostar diffusivity, the system was further equilibrated for $303.0\,\text{ns}$ at the target $\phi$.

To determine the diffusion coefficient $D$ of DNA nanostars at different $\phi$, MD simulations were performed for $2 \times 10^{9}$ time steps, corresponding to $\sim 30.3~\mu\text{s}$. The trajectory was sampled every $2 \times 10^{6}$ time steps, yielding 1,000 frames for analyzing the in-plane mean-square displacement of the nanostars' center-of-mass, $\langle \Delta r^2(t) \rangle$. The long-time self-diffusion coefficient of the nanostars was then determined by fitting $\langle \Delta r^{2}(t) \rangle = 4Dt$ (Fig. S8).
We have repeated the simulations three times, starting from different initial configurations, to improve the statistics.

\section*{Data availability}
The data that support the findings of this study are available from the corresponding authors upon reasonable request.

\section*{Author contributions}
K. M.: Conceptualization (equal); data curation: (equal); formal analysis (lead); investigation (equal); methodology (equal); validation (equal); visualization (lead); writing -- original draft (lead); writing -- review and editing (equal).
\\
T.Y.: Conceptualization (equal); data curation: (equal); formal analysis (supporting); investigation (equal); methodology (equal); software (lead); validation (equal); visualization (supporting); writing -- review and editing (supporting).
\\
M.Y.: Conceptualization (equal); funding acquisition (equal); project administration (equal); resources (equal); supervision (equal); writing -- original draft (supporting); writing -- review and editing (equal).
\\
A.N.: Conceptualization (equal); funding acquisition (equal); project administration (equal); resources (equal); supervision (equal); visualization (supporting); writing -- original draft (supporting); writing -- review and editing (equal).

\section*{Conflicts of interest}
There are no conflicts to declare.

\section*{Supporting information}
The Supporting Information is available free of charge at [URL]. Section S1: Experimental determination of nanostar localization using the relative fluorescence intensity, $I_{\mathrm{r}}$; Section S2: Estimation of the experimental packing fraction; Figure S1: DNA sequences of the rigid and soft nanostars; Figure S2: Representative fluorescence intensity profiles across the droplet center and schematic of the droplet cross section used to determine $I_{\mathrm{r}}$; Figure S3: Dependence of $I_{\mathrm{r}}$ on droplet radius; Figure S4: Representative FRAP recovery curves for freely diffusing nanostars in the droplet interior; Figure S5: Droplet-size dependence of the diffusion coefficient, $D$, and mobile fraction, $M_{\mathrm{f}}$, of DNA nanostars and lipids; Figure S6: Experimental FRAP results obtained at different positions on the droplets; Figure S7: Estimation of instantaneous area $S_0$ of DNA nanostars; Figure S8: Mean-square displacement (MSD) of rigid and soft nanostars at different $\phi$.

\section*{Acknowledgements}
We thank Takamasa Sakai for helpful discussions. This work was partially funded by the Japan Society for the Promotion of Science (JSPS) KAKENHI (grant numbers  22H01188, 24H02287 (M.Y.), the Japan Science and Technology Agency (JST) (grant numbers FOREST, JPMJFR213Y; CREST (JPMJCR22E1) (M. Y.)), the World-Leading Innovative Graduate Study Program for Advanced Basic Science Course (WINGS-ABC) at the University of Tokyo (K. M.), the Deutsche Forschungsgemeinschaft (DFG, German Research Foundation) through Project No. 470113688, and by the Klaus Tschira Foundation through Project No. 00.050.2024/ID 25347 (A. N.).


\bibliography{references.bib}

\end{document}


\renewcommand{\thesection}{S\arabic{section}}
\renewcommand{\thesubsection}{S\arabic{section}.\arabic{subsection}}
\renewcommand{\thefigure}{S\arabic{figure}}
\renewcommand{\thetable}{S\arabic{table}}
\renewcommand{\theequation}{S\arabic{equation}}

\maketitle

\noindent
Number of pages: 6 \\
Number of figures: 8

\tableofcontents

\clearpage
\newpage

\section{DNA sequences}
\begin{figure}[htb]
\centering
\includegraphics[width=1 \linewidth] {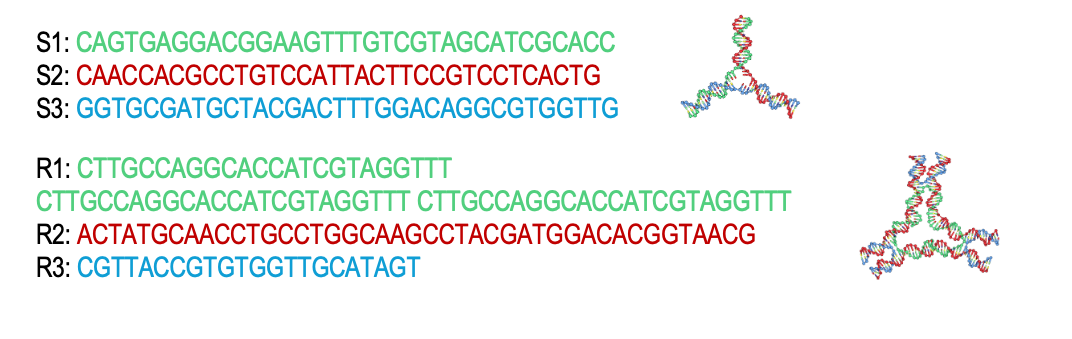}
\caption{
DNA sequences used to construct the DNA nanostars (S1-3 for a soft nanostar and R1-3 for a rigid nanostar). Each color corresponds to the same-colored strand shown in the simulation snapshot.
}
\label{fig:sequences} 
\end{figure}

\section{Estimation of the experimental $I_\mathrm{r}$ and its dependence on $1/R$}

\begin{figure}[htb]
\centering
\includegraphics[width=0.7 \linewidth] {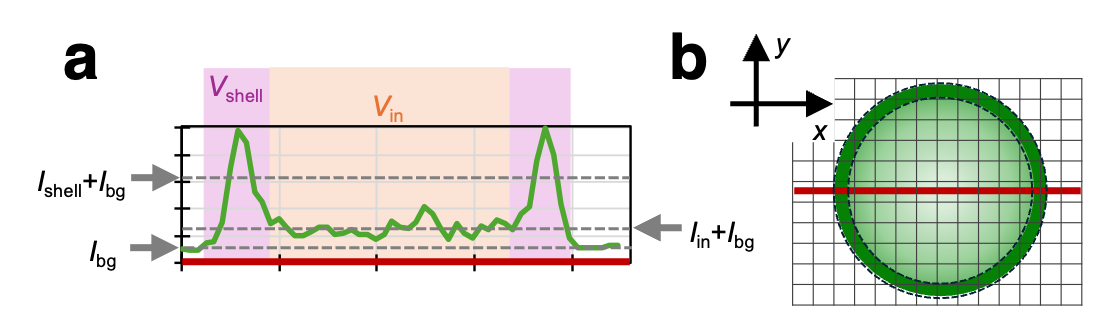}
\caption{
(a) Experimental determination of $I_\mathrm{r}$ from the fluorescence intensity profile of DNA nanostars. The mean background intensity, $I_{\mathrm{bg}}$, was first calculated, and the position where the intensity started to increase above $I_{\mathrm{bg}}$ was defined as the outer boundary of the droplet. The inner intensity, $I_{\mathrm{in}}$, was obtained by subtracting $I_{\mathrm{bg}}$ from the mean intensity inside the droplet, and the position where the intensity started to deviate from $I_{\mathrm{in}}$ was defined as the inner boundary of the DNA nanostar adsorption shell. The volumes of each region ($V_\mathrm{shell}$ and $V_\mathrm{in}$, respectively) were estimated from the corresponding radii. The volume-weighted mean intensity corresponds to the integrated fluorescence intensity divided by the total volume of the corresponding region.
(b) $I_\mathrm{all}$ and $I_\mathrm{in}$ were determined using ImageJ from pixel-resolved fluorescence intensity data in two-dimensional cross-sectional images. The mean pixel intensity within each region defined in (a) was used as $I_\mathrm{all}$ or $I_\mathrm{in}$.
}
\label{fig:schematic_intenisty_analysis} 
\end{figure}

Because not all encapsulated nanostars necessarily localize at the membrane interface, we introduced an experimentally determined localization factor, $I_\mathrm{r}$, which represents the fraction of the total fluorescence signal associated with the membrane-adjacent region. As illustrated in Fig.~\ref{fig:schematic_intenisty_analysis}, $I_\mathrm{r}$ was estimated from the fluorescence intensity profile together with the droplet geometry, and then used as a correction factor to experimentally estimate the interfacial packing fraction. The total fluorescence signal from a droplet can be expressed as the sum of the local fluorescence signals
\begin{equation}
I_\mathrm{all} V_\mathrm{all} = I_\mathrm{shell} V_\mathrm{shell} + I_\mathrm{in} V_\mathrm{in},
\end{equation}
where $I_\mathrm{all}$, $I_\mathrm{shell}$, and $I_\mathrm{in}$ are the mean fluorescence intensities of the whole droplet, the membrane-adjacent shell region, and the inner region, respectively. The corresponding volumes of those regions are
\begin{equation}
V_\mathrm{all}=\frac{4}{3}\pi R^3,
\quad
V_\mathrm{in}=\frac{4}{3}\pi R_\mathrm{in}^3,
\quad
V_\mathrm{shell}=\frac{4}{3}\pi\left(R^3-R_\mathrm{in}^3\right), 
\end{equation}
with $R$ and $R_\mathrm{in}$ being the radii of the whole droplet and the inner region, respectively. 

The membrane-associated fraction of fluorescence was then defined as
\begin{equation}
I_\mathrm{r} =
\frac{I_\mathrm{shell}V_\mathrm{shell}}
{I_\mathrm{all}V_\mathrm{all}} =
\frac{I_\mathrm{shell}V_\mathrm{shell}}
{I_\mathrm{shell}V_\mathrm{shell}+I_\mathrm{in}V_\mathrm{in}} =
\left(1+\frac{I_\mathrm{in}V_\mathrm{in}}{I_\mathrm{shell}V_\mathrm{shell}}\right)^{-1}.
\end{equation}

Since the confocal cross-section images are much thinner than the typical droplet radii, we neglect spherical curvature effects. Experimentally, $I_\mathrm{in}$ was obtained as the mean pixel intensity of the inner region in two-dimensional cross-sectional fluorescence images after background subtraction. Because the shell region is thin, $I_\mathrm{shell}$ was calculated from the mean intensity of the entire cross section, $I_\mathrm{all}$, and the mean intensity of the inner region, $I_\mathrm{in}$. Assuming that the fluorescence intensity is uniform within the shell and inner regions, respectively, the total fluorescence intensity in the cross-sectional image is given by
\begin{equation}
I_\mathrm{all} \pi R^2 =
I_\mathrm{shell}\pi\left(R^2-R_\mathrm{in}^2\right)
+ I_\mathrm{in}\pi R_\mathrm{in}^2 .
\end{equation}
Therefore, $I_\mathrm{shell}$ was calculated as
\begin{equation}
I_\mathrm{shell} =
\frac{I_\mathrm{all}R^2 - I_\mathrm{in}R_\mathrm{in}^2}
{R^2-R_\mathrm{in}^2}.
\end{equation}

\begin{figure}[htb]
\centering
\includegraphics[width=0.7 \linewidth] {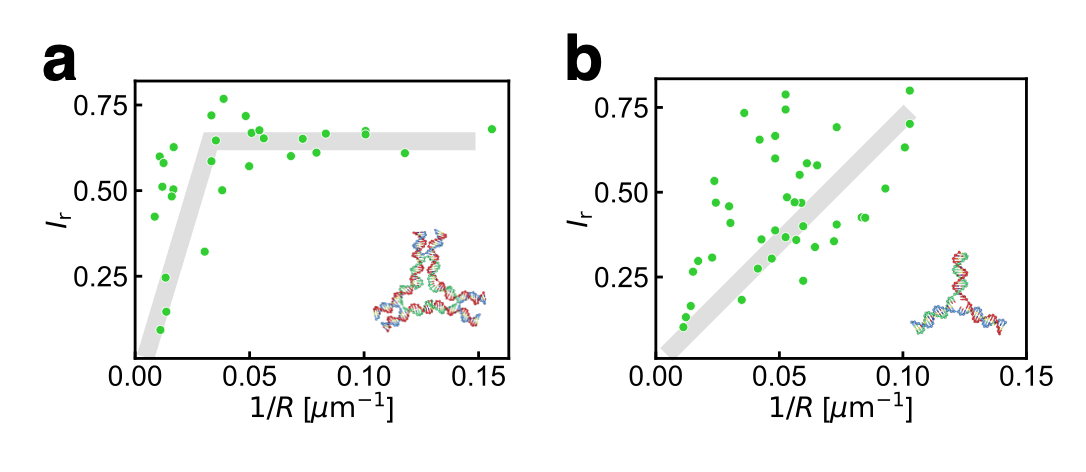}
\caption{
Localization factor $I_\text{r}$ determined for rigid (a) and soft nanostars (b) as functions of the inverse droplet radius $1/R$. The solid lines are guides to the eye only.
}
\label{fig:intensity_ratio} 
\end{figure}

For rigid nanostars, $I_\mathrm{r}$ increased approximately in proportion to $1/R$ only for large droplets and then reached a plateau at $1/R \gtrsim 0.03~\mu\mathrm{m}^{-1}$ ($R \lesssim 30~\mu\mathrm{m}$), indicating saturation of membrane localizaiton (Fig.~\ref{fig:intensity_ratio}a). In contrast, for soft nanostars, $I_\mathrm{r}$ increased almost linearly with $1/R$ over the entire droplet-size range examined, consistent with the expected surface-to-volume scaling of a spherical droplet (Fig.~\ref{fig:intensity_ratio}b). This difference is consistent with the larger molecular footprint, $S_0$, of the rigid nanostars, leading to membrane saturation in smaller numbers of nanostars than in soft nanostars.

\section{Estimation of the experimental $\phi$}
The experimental packing fraction of DNA nanostars on the droplet membrane, $\phi$ was estimated from $R$, $c_\mathrm{DNA}$, $S_0$ and $I_\mathrm{r}$. The number of membrane-associated nanostars was approximated as
\begin{equation}
N_\mathrm{mem}
=
I_\mathrm{r} c_\mathrm{DNA} N_\mathrm{A} V_\mathrm{all},
\end{equation}
where $N_\mathrm{A}$ is Avogadro's constant. The corresponding surface packing fraction was then calculated as
\begin{equation}
\phi
=
\frac{N_\mathrm{mem} S_0}{4\pi R^2}
=
\frac{I_\mathrm{r} c_\mathrm{DNA} N_\mathrm{A} R S_0}{3}.
\end{equation}
where $S_0$ is the projected area occupied by a single DNA nanostar on the membrane.
When $R$ is expressed in $\mu\mathrm{m}$, $S_0$ in $\mathrm{nm}^2$, and $c_\mathrm{DNA}$ in $\mu\mathrm{M}$, this expression becomes
\begin{equation}
\phi \approx \frac{2 R S_0 c_\mathrm{DNA} I_\mathrm{r}}{1000}.
\end{equation}

\section{Additional experimental results}

\begin{figure}[t]
\centering
\includegraphics[width=0.7 \linewidth] {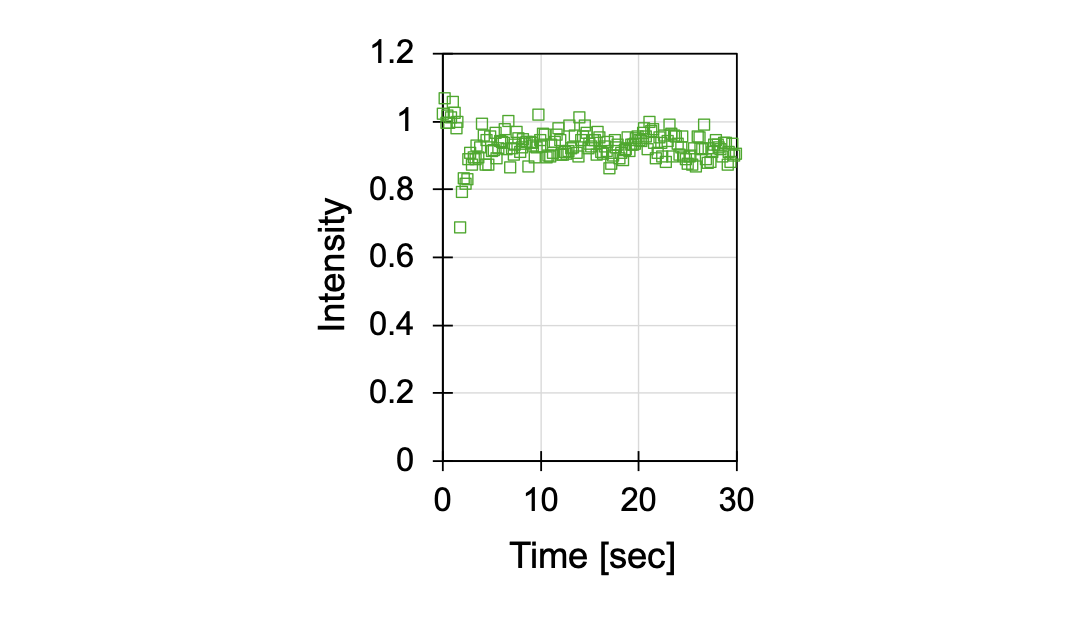}
\caption{
A representative FRAP curve for inner rigid nanostars freely diffusing in the interior of large droplets with $R \approx 90\,\mu\text{m}$ at $c_{\mathrm{DNA}} = 2.5~\mu\text{M}$.
}
\label{fig:frap_inside} 
\end{figure}

\begin{figure}[t]
\centering
\includegraphics[width=1.0 \linewidth] {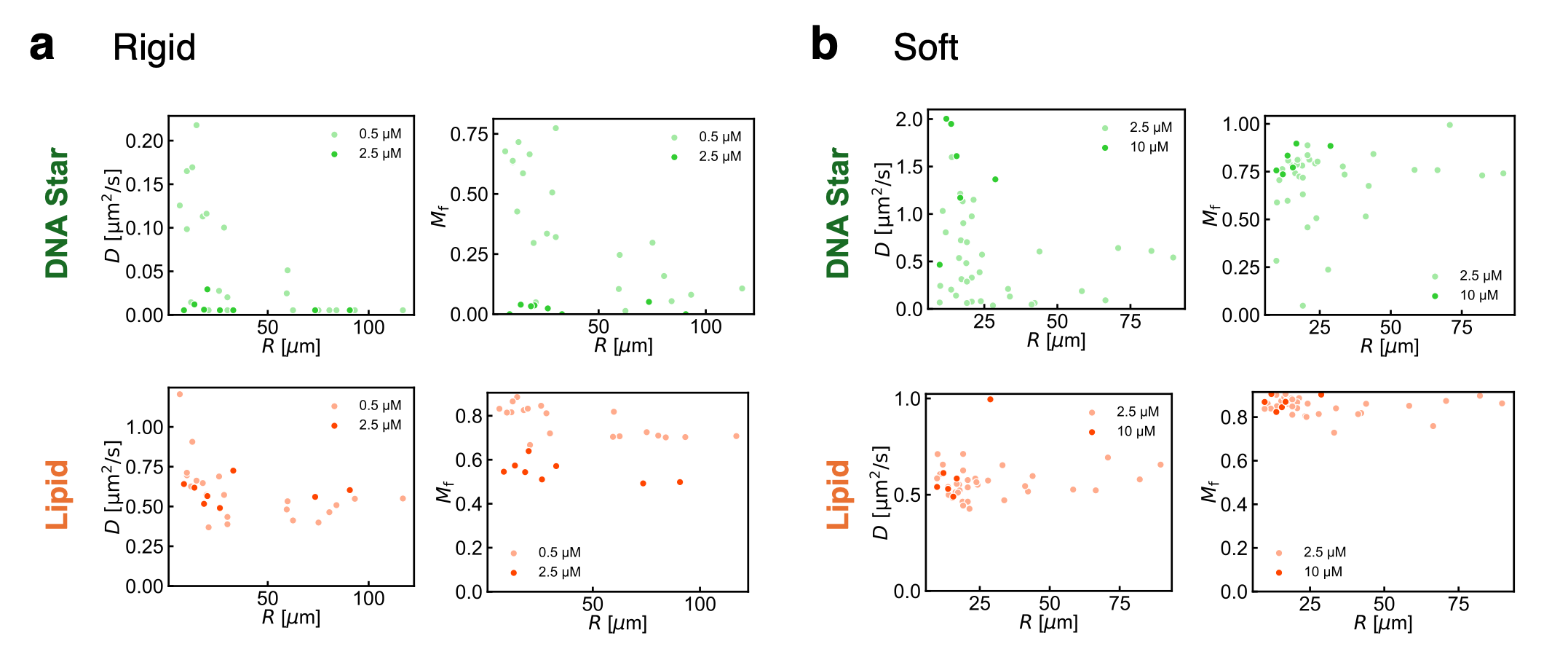}
\caption{
Diffusion coefficients $D$ and mobile fractions $M_\mathrm{f}$ of (a) lipids and (b) DNA nanostars on DOTAP membranes covering droplets with varying radii, $R$, measured for two DNA concentrations ($c_\mathrm{DNA}$ = 0.5 and 2.5 $\mu$M). 
}
\label{fig:frap_original} 
\end{figure}

\begin{figure}[t]
\centering
\includegraphics[width=0.7 \linewidth] {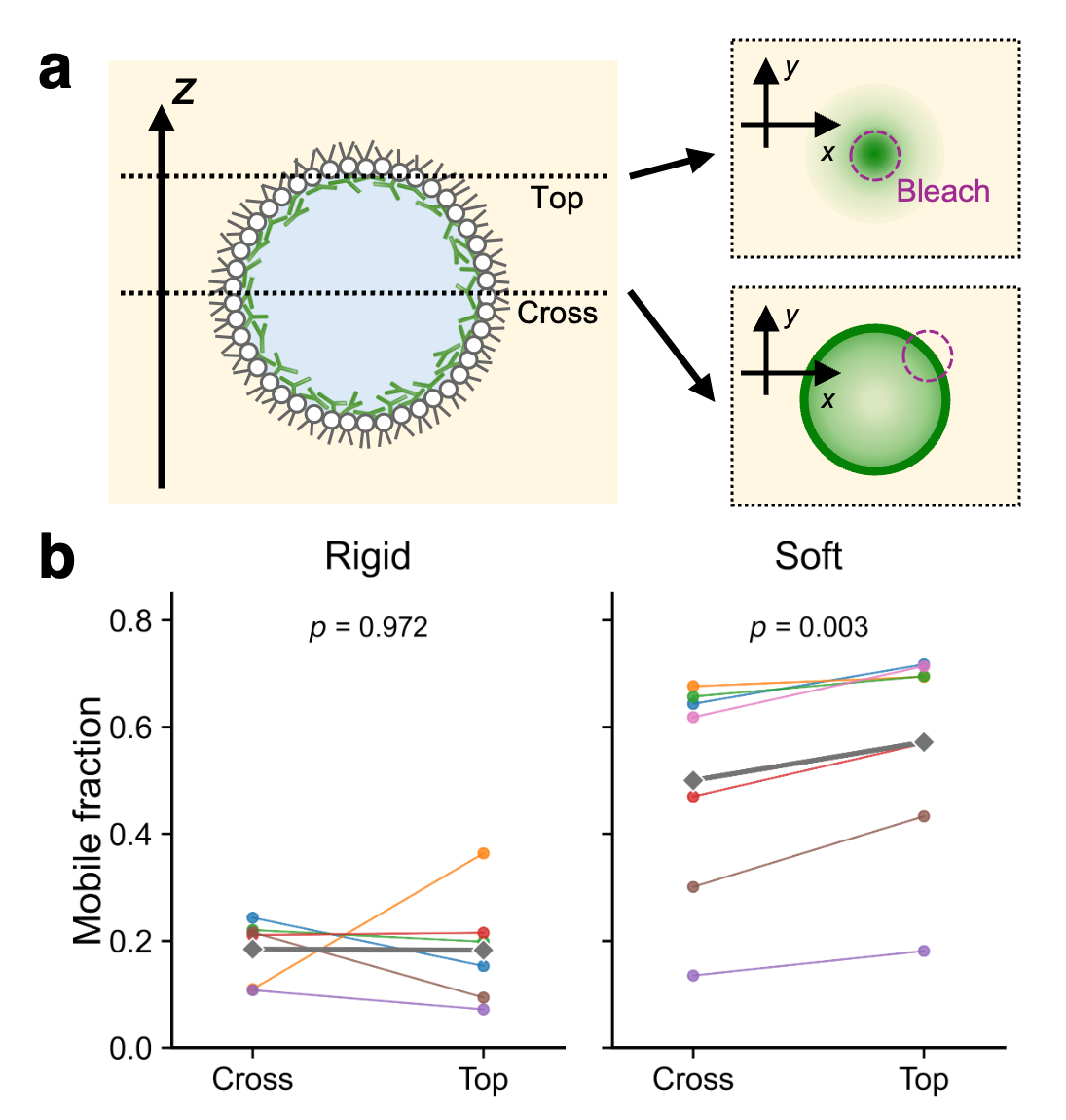}
\caption{(a) Schematics of FRAP experiments conducted at the top and cross-sectional regions of the droplet. Fluorescence recovery is dominated by reversible adsorption–desorption at the top and by lateral diffusion in the cross-sectional region.
(b) Mobile fraction of rigid and soft nanostars at the top and cross-sectional regions. Statistical significance between the cross-sectional and top measurements from the same droplets was assessed using a two-sided paired $t$-test ($p$ values).
}
\label{fig:frap_diff_loc} 
\end{figure}

\clearpage
\section{Estimation of instantaneous area of DNA nanostars}
\begin{figure}[htb]
\centering
\includegraphics[width=1.0 \linewidth] {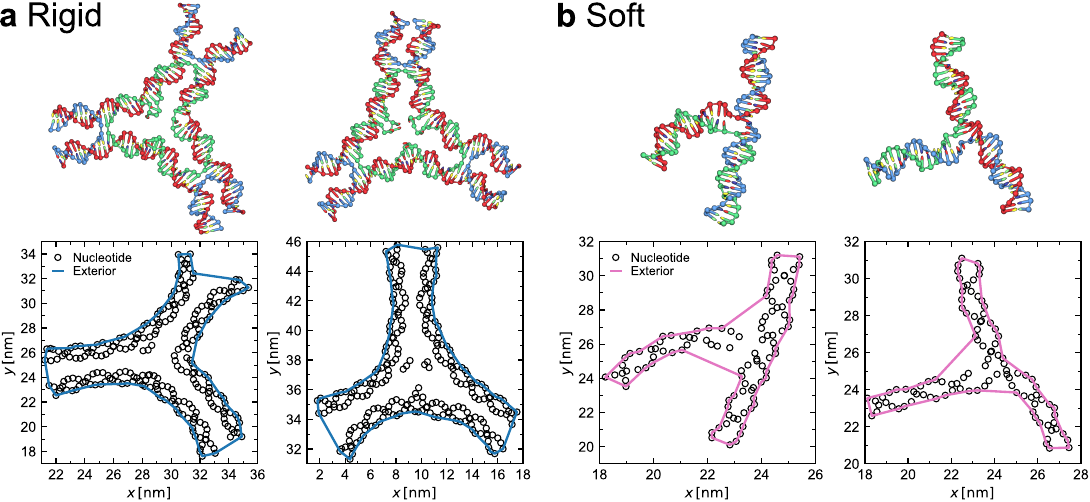}
\caption{Representative DNA nanostar configurations and their projections onto the $xy$-plane (black points) for (a) rigid and (b) soft nanostars. The blue and pink lines indicate the outer boundaries of the nanostars, and $S_0$ is defined as the area enclosed by these boundaries.}
\label{fig:S0} 
\end{figure}

We characterized the size of the DNA nanostars by their instantaneous area, $S_0$, defined determined from single-nanostar simulations. 
The value of $S_0$ was obtained by projecting the nucleotide positions onto the $xy$-plane, identifying the exterior nucleotides by Delaunay tessellation on the projected coordinates (Fig.\ref{fig:S0}), and then calculating the area enclosed by the resulting nanostar boundary.
We took the average over $10^4$ different configurations for both rigid and soft DNA nanostars.

\section{Mean-square displacement of DNA nanostars from MD simulations}
\begin{figure}[htb]
\centering
\includegraphics[width=1.0 \linewidth] {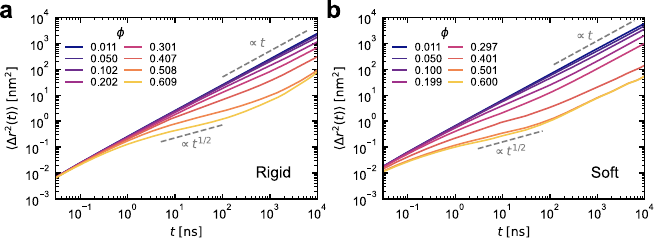}
\caption{In-plane mean-square displacement $\langle \Delta r^{2} (t) \rangle$ of the center of mass of (a) rigid and (b) soft DNA nanostars at different $\phi$.}
\label{fig:MSD} 
\end{figure}